\documentstyle[11pt,aas2pp4]{article}
\begin{document}

\title{Automatic Redshift Determination
by use of Principal Component Analysis --- I: Fundamentals}

\author{Karl Glazebrook, Alison R.Offer and Kathryn Deeley}
\affil{Anglo-Australian Observatory}

\authoraddr{Anglo-Australian Telescope, Siding Spring Observatory, Coonabarabran, 
NSW 2357, AUSTRALIA}

%\date{Accepted 1995 May 17. Received 1995 March 29}
%\pubyear{1997}

%%%%%%%%%%%%%% KGB's misc macros %%%%%%%%%%%%%%

\def\Msun{\hbox{$M_{\odot}$}}
\def\Hunits{\hbox{$\rm km\,s^{-1}\,Mpc^{-1}$}}
\def\phiunits{\hbox{$h^3\,\rm \hbox{Mpc}^{-3}$}}
\def\hkpc{\hbox{$h^{-1}\,\rm \hbox{kpc}$}}
\def\hMpc{\hbox{$h^{-1}\,\rm \hbox{Mpc}$}}
\def\ie{\hbox{\it i.e.}}
\def\HST{{\it HST\/}}
\def\etal{{\it et~al.}}
\def\gs{\mathrel{\lower0.6ex\hbox{$\buildrel {\textstyle >}
 \over {\scriptstyle \sim}$}}}
\def\ls{\mathrel{\lower0.6ex\hbox{$\buildrel {\textstyle <}
 \over {\scriptstyle \sim}$}}}
\def\micron{\hbox{$\mu\rm m$}}
\def\sbunits{\hbox{$\hbox{mags}/\hbox{arcsec}^{-2}$}}

\let\ds=\displaystyle
\def\wl{\hbox{$w_{\lower1pt\hbox{$\scriptstyle \lambda$}}$}}

%%%%%%%%%%%%%%%%%%%%%%%%%%%%%%%%%%%%%%%%%%%%%%

\label{firstpage}

%\maketitle

\begin{abstract}
With the advent of very large redshift surveys of tens to hundreds
of thousands of galaxies reliable techniques for automatically 
determining galaxy redshifts are becoming increasingly important. The
most common technique currently in common use is the cross-correlation
of a galactic spectrum with a set of templates. This series of papers
presents a new method based on Principal Component Analysis. The  method
generalizes the cross-correlation approach by replacing the individual
templates by a simultaneous linear combination of orthogonal templates.  
This effectively eliminates the mismatch between templates and data  
and provides for the possibility of better error estimates.  In this 
paper, the first of a series, the basic mathematics are presented along with a simple 
demonstration of the application.

\bigskip
\bigskip
\begin{center}{\it Submitted to Astrophysical Journal}\end{center}
\end{abstract}


\section{INTRODUCTION}

The development of fiber-based spectrographs capable of observing hundreds
of objects simultaneously has led to the advent of many large redshift
surveys with the intention of furthering our understanding of the large
scale structure, clustering and evolution of galaxies.  (For a review
see Strauss 1996.)
Examples include the Las Campanas Redshift Survey of 26,000
galaxies, just completed (Shectman \etal\ 1996), and the Two Degree Field (2dF)
Redshift Survey (Taylor \etal\ 1997, Maddox \etal\ 1997) 
which has started this year (1997)
and will measure the redshifts of over 250,000 galaxies over the next
several years.

Because of the sheer size of these surveys it is becoming very important
to develop methods of reliably, and quantifiably, measuring the redshifts
of the galaxy spectra without manual intervention. For example in the 2dF
survey a method with a 95\% success rate would still leave 12,500 spectra
to be inspected manually, a very large task. Ideally any automatic redshift
calculation should also give an accurate error estimate and confidence
rating for each redshift to indicate {\em which\/} 12,500 galaxies 
out of the 250,000 need further, possibly manual, attention.

At the current time the most successful and widespread method of 
automatic redshift measurement is cross-correlation
analysis (\cite{td}).  In this method the galaxy 
spectrum is cross-correlated  with a series of template spectra
corresponding to a sequence of standard galaxy or stellar types.  
The size of the largest peak in the cross-correlation function
is an indication of the quality of the match between
the galaxy and the template spectrum. The position and width of the peak
give the redshift and an `error' on the redshift. If the galactic
and template spectra agreed exactly then a sharp correlation
peak would be found, but in practice it is unlikely that the
galactic spectrum will exactly match \em any \em of the
template spectra. Depending on the size of the mismatch, the
redshift may or may not be correct - the `error' is merely a measure of
the accuracy of the location of the peak and not an indication of its
true worth.  Tonry \& Davis 
presented a formulation for the error
on peak location, which was improved upon by Heavens (1993).

A series of templates consisting of different types of galactic
spectra, individually tested, is not necessarily the optimal template
set to use. It would be preferable to generalize the concept of
cross-correlation to use a simultaneous linear combination of templates,
with expansion coefficients that depend on the redshift. With a suitable
choice of template spectra, the mismatch between the data and a linear
combination of a small number of template spectra could be reduced to 
an arbitrarily small amount. Any residual would be due only to the
random component of the observational noise. 

In this paper, the first of a series of papers,
a method is presented for achieving this. The method, which we will
call `PCAZ', is  based upon the use of Principal Components Analysis
to make the general linear problem amenable to efficient computation.
The fundamental mathematics is presented in section 2, and a simple
demonstration based upon some sample 2dF galaxy spectra is shown in section 3. 
Subsequent papers will present in more detail the methods 
of robust error analyses and software for implementing the PCAZ 
algorithm.

\section{MATHEMATICS BEHIND PCAZ}

\subsection{Standard Cross-Correlation Revisited}

Consider a galaxy spectrum $G_\lambda$ (with normally
distributed errors, variance $\sigma^2_\lambda$) 
requiring a redshift $z$, and a single template spectrum $T_\lambda$. If
both the galactic spectrum and the template spectrum are
binned on the same logarithmic wavelength grid the
likelihood that the galaxy and the template are the same, bar the
redshift and normalization, can be written:

\begin{equation}
\label{like}
-\log {\cal L} \propto
\chi^2 = \sum_{\lambda} { {1\over \sigma^2_\lambda} \left[ G_\lambda - 
                 a(z) \, T_{\lambda+\Delta} \right]^2}
\end{equation}
where the sum is over discrete wavelength
bins, $(\lambda=1,2,3,\dots)$ and $\Delta$ is the linear
shift along the logarithmic grid due to the redshift,
$\Delta  \propto \log(1+z)$. $a(z)$ is the
redshift dependent coefficient of the template. At any particular redshift
$z$ we can find the value of $a(z)$ that maximizes the likelihood ({\it i.e.}
gives the best match between the galaxy and the template) by
setting $\partial \chi^2 /\partial a = 0$. 
Solving this gives:
\begin{equation}
\label{a(z)}
 a(z) =  {
           \ds\sum_{\lambda}   
            {\ds 1\over \ds \sigma^2_\lambda} \ G_\lambda T_{\lambda+\Delta} 
           \over
             \ds\sum_{\lambda} {\ds 1\over \ds \sigma^2_\lambda} \ T_{\lambda+\Delta}^2
           }
\end{equation}

It can be seen that $a(z)$ in equation~\ref{a(z)} is simply proportional to the 
cross-correlation function of the  galaxy spectrum with the template
spectrum, for the case where the variance is ignored.
Substituting this value of $a(z)$ 
into equation~\ref{like} and simplifying gives:
\begin{equation}
 \label{finalchi2}
 \chi^2 = \left(\sum_{\lambda}   {1\over \sigma^2_\lambda} {G_\lambda^2 }
          - a(z)^2 \left( \ds\sum_{\lambda}  {1\over \sigma^2_\lambda} {  
          T_{\lambda+\Delta}^2} \right)^{-1}
          \right)
\end{equation}
The minimum of $\chi^2$ as a function of redshift
occurs when $a(z)$ is maximized,
thus finding the peak of the cross-correlation function (or CCF) is 
exactly equivalent to finding the maximum likelihood redshift at 
which the single template best matches the data (again putting in the
variance introduces complications, see section\ref{varsec}).

This maximum likelihood basis for cross-\-correl\-ation is fundamental to the linear
generalization, but has not been remarked upon in the astronomical
literature. The approach that has been used historically
to assign a confidence or quality value to the redshift has been based upon
the height of the CCF peak above the CCF `noise' (see for example
\cite{heav}). However much of this `noise' is due to systematic mismatch
between the template and the data rather than observational noise
and thus the assumption that peaks are uncorrelated
is invalid. With the formulation given above, and realistic errors, 
it would be possible to assign a true likelihood value and hence confidence intervals
to a peak {\em if} the 
template and the galaxy were identical and just differed due to the observational
noise and the redshift.

\subsection{Linear Generalisation}

The standard cross-correlation method of Tonry \& Davis tests the candidate galaxy spectrum
against a range of template spectra individually. The linear generalization presented
here essentially assumes that a galaxy spectra can be expanded as a linear sum of 
template spectra. This in principle allows the systematic mismatch 
between galaxy and template to be arbitrarily reduced and hence a realistic likelihood
to be assigned to an output redshift.

Initially, for simplicity, we will consider how one solves for the
values of the coefficients at zero redshift. We assume a galaxy
spectrum is represented by an $n$ dimensional vector $G$, where $n$ is the number 
of wavelength bins. The $m$ template spectra are represented by the rows of an $m\times n$
matrix $\bf T$. The galaxy spectrum is then fitted by a linear combination of templates
with coefficients $a_j$:
\begin{equation}
 G_\lambda \simeq \sum_{j} a_j T_{j\lambda} 
\end{equation}
The coefficients, $a_j$, may be found by following the same maximum likelihood
recipe used above and minimizing $\chi^2$  where now:
\begin{equation}
\chi^2 = 
\sum_\lambda {\wl^2\over \sigma^2_\lambda} \left[G_\lambda-\sum_j {a_j T_{j\lambda}}\right]^2 
\end{equation}
We have now introduced  $\wl$ as representing a general wavelength
dependent weighting function (which might be used, for example, to
emphasize particular spectral features). Setting $\partial \chi^2/\partial a_i = 0$ 
leads to the matrix equation:
\begin{equation}
\label{CA=TG}
{\bf C a} = {\bf T G'} 
\end{equation}
where the elements of the vector $\bf G'$ are $G'_{\lambda}=\wl^2 G_{\lambda}/
\sigma^2_\lambda$
and the elements of the $m\times m$ correlation matrix, {\bf C}, are given by:
\begin{equation}
\label{CWTT}
C_{ij}={\wl^2\over \sigma^2_\lambda} T_{i\lambda} T_{j\lambda}
\end{equation}

Direct inversion of the $\bf C$ matrix to obtain the $a_j$ coefficients
is clearly impractical. Not only would it be numerically intensive to do
this at many trial redshifts,
but the presence of very small eigenvalues (see section~2.2 below) 
would lead to large numerical instabilities. However, if the template spectra 
(the rows of the matrix $\bf T$) are replaced by a basis set of 
orthogonal vectors, the transformed correlation matrix will be diagonal, 
and the problem simplifies.  Principal Component Analysis (PCA) is 
the tool used to select the orthogonal vectors.

\subsection{Principal Component Analysis}

Principal Component Analysis is a technique frequently used
for data compression and classification (\cite{advstat}, \cite{murtagh}). 
In particular, direct PCA of spectral data similar to that used 
here has been used for classification of galactic spectra 
(\cite{mittaz,cea,fea,sodre}) and for classification 
of QSO spectra (\cite{francis}).

In essence, PCA finds the `best' representation of a set of data by 
a set of orthogonal vectors, or principal components, which
can be combined linearly to reconstitute the data. The components
are ordered in terms of significance in a least squares sense
and data compression is achieved by retaining only the most significant
principal components. 

PCA can be formulated in two different but equivalent ways, both
of which have been used for spectral classification. Consider a
set of $m$ template spectra sampled at $n$ discrete wavelengths.
The elements of the matrix $\bf T$ can be pictured  as a series of
row vectors, each of which is a point representing a spectrum
in $n$-dimensional wavelength space. Alternatively, the data can
be thought of as column vectors with
each point in $m$-dimensional template space 
being the set of fluxes in an individual wavelength bin.
A PCA in the template space diagonalizes 
the elements of the $m\times m$ correlation matrix:
\begin{equation}
\label{CTT}
%
% C_{ij}=  \sum_{\lambda}\wl^2 T_{i\lambda} T_{j\lambda}
%
C_{ij}=  \sum_{\lambda} T_{i\lambda} T_{j\lambda}
\end{equation}
where we, for now, ignore weights and variance factors
for clarity in the discussion.
A PCA in wavelength space diagonalizes the elements of the 
$n\times n$ correlation matrix:
\begin{equation}
D_{\lambda_1\lambda_2}=  \sum_{i} T_{i\lambda_1} T_{i\lambda_2}
\end{equation}
The two approaches are equivalent and in principle they will 
lead to the same eigenvalues and principal components that
are related by a simple transformation (\cite{murtagh}). In many ways
the wavelength space is more intuitive for spectral classification
and Mittaz \etal\ (1990), Francis \etal (1992), Folkes \etal\ (1996)
and Sodr\'e \& Cuevas (1997) have 
used a PCA in wavelength space for this. 
Connolly \etal\ (1995), who only used a small number of 
spectra, chose to work with the reduced dimensionality of template space.
In order to clearly show the link between the cross-correlation
method and PCA, and because we have fewer template spectra 
than wavelength bins we have chosen to follow 
Connolly \etal\ and 
perform the diagonalization in template space.

In practice this means taking the set of templates and constructing
from them a set of orthogonal `eigentemplates'. The
matrix $C$ is diagonalized, to yield a set of eigenvalues:
\begin{equation}
 C = R \Lambda R^T \hbox{\quad where\ \ } \Lambda = \left(\matrix{ 
\Lambda_1 & 0         & 0         & \dots \cr
0         & \Lambda_2 & 0         & \dots \cr
0         & 0         & \Lambda_3 & \dots \cr
\vdots    & \vdots    & \vdots    & \cr
}\right) 
\end{equation}
and associated matrix of $m$-dimensional eigenvectors, $\bf R$, which
are the principal components in template space. 
The diagonalization is accomplished by standard numerical techniques
such as Singular Value Decomposition (\cite{mittaz}). The matrix $R$
defines a transformation  between the 
template spectra and a set of $n$-dimensional {\em orthogonal} 
eigentemplates, the principal components in wavelength space $E_{i\lambda}$:
\begin{equation}
E_{i\lambda} = \sum_k R_{ki} T_{k\lambda}  
\end{equation} 
This is essentially the Karhumen-Lo\`eve transform (\cite{murtagh}). The
resulting eigentemplates satisfy the orthogonality property:
\begin{equation}
\label{EELD}
%
% \sum_\lambda \wl^2 E_{i\lambda} E_{j\lambda} = \Lambda_{i}\delta_{ij}
%
  \sum_\lambda E_{i\lambda} E_{j\lambda} = \Lambda_{i}\delta_{ij}
\end{equation}
where $\delta_{ij}$ is the Kronecker--$\delta$.  The eigenvalues
$\Lambda_{i}$ represent the contribution of each eigentemplate to the 
set of templates in a least squares sense.  If the principal components
are arranged in order of decreasing eigenvalue it can be shown (\cite{advstat})
that the first principal component in either space is the line along which
the cloud of points is the most elongated (has the greatest variance).
Equivalently, the first principal component is the line for which the 
sum of the squared perpendicular distances
of the points from the line is a minimum. Similarly, if the points are
projected onto a hyperplane orthogonal to the first principal component,
the second principal component is the line in that hyperplane along which the
projected distribution is most elongated. Representing the data in terms of
just the first principal component would be equivalent to approximating the 
cloud of points by a line and characterizing each point in terms of its projected
distance along the line. Representing the data in terms of the first two
principal components is equivalent to projecting the cloud of points onto a
plane. 

The spectra within the template set can be represented to any 
given accuracy by a linear combination of eigentemplates:
\begin{equation}
\label{reconst}
 G_{\lambda} \simeq \sum_{j=1}^{p} b_j E_{j\lambda} 
\end{equation}
where $p$ is the number of eigentemplates retained. 
Since the eigentemplates are orthogonal the
corresponding expansion coefficients are given by:
\begin{equation}
\label{bGEL}
%
% b_j = {\wl^2 G_{\lambda} E_{j\lambda} \over \Lambda_j} 
%
 b_j = \sum_{\lambda} { G_{\lambda} E_{j\lambda} \over \Lambda_j} 
\end{equation}
where $\Lambda_j$ is the $j$th eigenvalue derived above. 

In practice only a subset of the principal components represent
real correlations and anticorrelations between the spectra within
the template set. The remaining principal components may contain a
large fraction of uncorrelated noise in which case they can 
be discarded. Folkes
\etal\  show that the number of significant
principal components, $p$, depends on the quality of the template data
set. Reconstruction of the template spectra from the first $p$
principal components effectively filters out much of the noise.
In the case where the input template set consists of 
a few very high signal/noise
spectra it may be desirable to retain all the eigentemplates --- 
in this case the PCA analysis can be viewed as a shortcut for
speeding up the solving of equation~\ref{CA=TG} for a large number
of redshifts.

To apply PCA to redshift determination it is necessary to
assume that the template set is sufficiently general that any galactic 
spectrum not included in the original template set 
can also be represented to the required accuracy by a summation over the first
$p$ principal components. Essentially we are assuming that the
correlations within the template set reflect a global correlation across
all galaxies in the survey. Allowance for abnormal objects such as stars,
active galaxies and quasars can be made by including example spectra
of these in the template set or by discriminating against bad matches
(see Section~3.2).

\subsection{Relation to Cross-Correlation and Redshift Determination}
\label{varsec}

The discussion of PCA above is general and up to this point follows
the spirit of the spectral classification of Mittaz \etal,
Francis \etal, Connolly \etal\ 
and Folkes \etal. The extra step is to include
the  redshift $z$ as an additional variable. Weighting can be retained, but
must be tied to the {\em rest frame\/} of the templates, the variance must
be assumed wavelength independent (but see below).
The 
coefficients of the eigentemplates then become:
\begin{equation} 
\label{bjdef}
b_j(z) = \sum_\lambda {w_{\lambda+\Delta}^2 G_\lambda E_{j(\lambda+\Delta)} \over \Lambda_j} 
\end{equation}
where each $b_j(z)$ is the cross-correlation function of the galaxy spectrum with
the $j$th eigentemplate weighted by the corresponding eigenvalue. 
If $g_k$ is the discrete fourier transform of the galaxy spectra,
$G_\lambda$ and $e_{jk}$ is the discrete fourier transform of 
$w_\lambda^2 E_{j\lambda}$ then the coefficients $b_j(z)$ are given 
by the inverse fourier transform of the product of $g_k$ and $e_{jk}$.

\begin{equation}
b_j(z) = {1\over N\Lambda_j} \sum_{k=0}^{N-1} g_k e_{jk} 
\exp{\left({2\pi ik\Delta\over N}\right)}
\end{equation}

The orthogonality gives the simple relation for the
joint likelihood:
\begin{equation} 
\label{newchi2}
-\log {\cal L} \propto
\chi^2 = {1\over\sigma^2} \left( \sum_\lambda w_{\lambda+\Delta}^2 
G_\lambda^2 - \sum_j \Lambda_j  b_j^2(z) \right)
\end{equation}
where the variance is $\sigma^2$.
The minimum of equation~\ref{newchi2} gives the maximum likelihood redshift,
$z$ through:
\begin{equation}
z_{\Delta_{min}}= 10^{\Delta_{min}\delta(\log\lambda)} - 1.
\end{equation}
where $\Delta_{min}$ is the shift that gives the minimum value 
of $\chi^2$. Note that the single cross-correlation function 
in equation~\ref{finalchi2} has been replaced in equation~\ref{newchi2} by a weighted sum 
of the squares of the individual  cross-correlation functions. 
This is a natural result given that the eigentemplates are orthogonal.

To preserve the orthogonality of templates with redshift it is not 
possible to weight by
a wavelength dependent variance, because the errors will be tied to the
observed frame. 
For a strongly wavelength dependent variance the method still gives the 
optimal fit in a least squares sense by minimizing the function:
\begin{equation} 
\label{neweqn}
f =   \sum_\lambda ( G_\lambda - \sum_i   b_i(z) E_{i,\lambda+\Delta})^2 
\end{equation}
One can then reintroduce $\sigma_\lambda$ for a final pass and calculate
a true likelihood for the final redshift, though this may not be the
absolute maximum likelihood because it is not used in the calculation of
the $b_j$'s. In practice this will mean a few more objects may fail to
have their redshift determined within specified likelihood bounds.

It should be noted that 
the logarithmic wavelength scale used for redshift
determination gives the correct weighting of spectral features. Since
$\Delta\lambda <<\lambda$ then
$\log(\lambda+\Delta\lambda)-\log(\lambda)\approx 
\Delta\lambda \log(\rm e)/\lambda$ so the fractional wavelength 
range per bin is constant across the spectrum. The important features 
in the eigenfunctions are the spectral lines which 
should be equally weighted. For a classical grating,
$\Delta\lambda\propto \lambda$ for unresolved lines, and for Doppler
broadening the same holds true. Thus logarithmic binning gives correct
equal weighting of features. Of course most real systems give close
to a linear wavelength scale, so the spectra must be resampled to
logarithmic bins which will introduce covariance between neighboring
pixels. However this will be very small as the wavelength scale only
changes very slowly across the spectrum. 

The PCAZ method has numerous advantages over previous methods in the literature:
\begin{enumerate}
\item Because it is just a set of cross-correlations the standard 
Fast Fourier Transform (FFT) method can be used to efficiently compute the $b_j(z)$'s. The 
simultaneous combination of $m$ eigentemplates takes the same computer
time as doing $m$ templates separately.
 
\item Existing cross-correlations codes can be used with little modification. They
only need to be provided with orthogonalized eigentemplates instead of the normal
templates as inputs and have some provision made for combining the
cross-correlation functions in quadrature afterwards.

\item Emission line galaxies are easily handled by PCAZ. 
The standard cross-correlation method gives relatively poor
results for these because emission line ratios vary much more than
absorption lines and hence
can no longer be accounted for by a small number of standard galaxy spectra. 
However, with the extra freedom given by a linear combination 
of eigentemplates variable line ratios can be accommodated. This
freedom means the method is robust against other wavelength-dependent
variations such as
only very approximate, or the absence of, flux calibration of the 
input spectra.

\item High signal/noise ($S/N$) eigenspectra can be 
created from a large set of noisy data as well 
as a small set of high $S/N$ spectra because each eigenspectra
represent an average of that mode over the data.
This would be especially suitable for a deep redshift survey 
where many of the weak ultraviolet absorption features would be missing
from local templates. A few hundred high redshift galaxies could have their 
redshifts measured manually. Eigenspectra constructed from these could be
used to measure the rest automatically.

\item The ability to calculate a likelihood means a true confidence could be
assigned to a redshift and future science analyses of survey statistics such as the
power spectrum of galaxy clustering $P(k)$ could include a realistic probability
distribution of redshift errors rather than neglecting them.
This is especially important for the next generation of very large surveys.

\item The maximum likelihood reconstruction from the coefficients $b_j E_{j\lambda}$ is a
noise filtered version of the data, which is useful for other analyses.

\item The coefficients $b_j$ are have independent errors and could be
used as the basis for classification scheme for faint spectra, either
by themselves or as input into other systems such as Artificial Neural Net
algorithms (e.g. \cite{fea}).

\item 
The provision of weights allows templates to be defined only in
regions of interest, for example around strong lines. This would,
for example, be particularly suitable
for very faint low $S/N$ data where one might wish to search for weak emission
lines appearing above the noise. With weights the rest of the noisy, possibly
undetected, continuum can be excluded from the $\chi^2$.

\end{enumerate}

\subsection{Practicalities}

There are a number of important practicalities involved in using the
PCA formalism to determine redshifts. The first is the issue of mean
subtraction. It is usual in PCA to subtract the mean of the distribution
from each point, in the case of spectral classification the mean spectrum
is subtacted from each of the template spectra prior to orthogonalization.
This is equivalent to moving the origin of the PCA co-ordinate system to
the center of the distribution of points. However, strictly a redshifted
mean spectrum should also be subtracted from the candidate spectrum whose
redshift is not yet known. Because of this, the mean spectrum was not
subtracted prior to orthogonalization.  

A second important point is continuum subtraction. Spectral classification
schemes have avoided continuum subtraction in order to retain as much
spectral information as possible (\cite{mittaz,cea,fea}). However, 
continuum subtaction
is more important for redshift determination. Continuum subtraction 
reduces the smoothly  varying background to zero and essentially
has the same effect as filtering out the long period fourier components
of the spectra. Without continuum subtraction the cross-correlation functions
show a broad peak representing the cross-correlation of the two apodized
continua, with a small spectral cross-correlation peak superimposed.

A final practicality is the normalization of the template spectra. Francis \etal, 
Folkes \etal\ and Sodr\'e \&
Cuevas normalize to unit flux:
\begin{equation}
\sum_\lambda T_\lambda = 1
\end{equation}
The alternative is to normalize to unit scalar product (\cite{cea}):
\begin{equation}
\sum_\lambda T_\lambda^2 = 1
\end{equation}
With continuum subtraction the resulting spectra 
oscillate about zero so normalization to unit scalar product was used.

\section{EXAMPLES}

In this section the method is illustrated using a set of sample
sky subtracted spectra. The method was developed using test spectra
from a variety of sources, we choose to illustrate its effectiveness
here using some early data recently taken from the 2dF galaxy survey for
which the algorithm is being developed. The 2dF survey is more comprehensively described
elsewhere: see Taylor \etal\ (1994, 1997) for a description of the 2dF instrument and 
Maddox \etal\ (1997) for an introduction to the galaxy survey.
The data described here consist of two test fields, SGP463 and NGP359, 
taken during 2dF commissioning for the survey in January--April 1997. The galaxies are selected
from the APM survey (\cite{maddogs-and-englishmen}) with $b_j<19.7$.

The 2dF spectra spanned a wavelength range of 3810\AA\ to 8227\AA\ with a 2 pixel
resolution of around 8.4\AA\ (line full-width half-max). 
Two fields were considered, one as template spectra and one as candidate
spectra whose redshift was to be determined. Redshifts had been previously assigned to this
both data sets by visual inspection
(M. Colless and K. Glazebrook private communication). 
This gives a typical accuracy of $\Delta z \simeq 0.0005$
set by the spectral resolution. The template field
contained a total of 91 galaxies for which redshifts had been assigned and the candidate field
contained 104 galaxies with known redshifts. The typical signal/noise of the
continuum was 10--30 at 5500\AA\ which should be typical for the survey spectra.
These are quite high signal/noise and we expect a variety of methods to work well,  in
the analysis below we add artificial noise to degrade the spectra to test robustness
of the method.

\subsection{Eigenspectra}

Two sets of eigenspectra were constructed. The first used  
five high signal/noise template spectra taken from an 
atlas of integrated spectra of local galaxies (\cite{kenni}). 
The five spectra chosen are listed in Table \ref{kensp}. They
cover a wavelength range of 3600\AA\ up to 7050\AA.
The spectra were rebinned on  a log wavelength grid with a grid spacing of 
$\delta\log_{10}(\lambda/\rm\AA)=1.7\times 10^{-4}$. 
The second set of spectra were derived from the 2dF data itself.  The
NGP359 field was used. The 91 spectra with well determined redshifts were 
corrected for redshift and used as the template set. A 
wavelength grid of 3100\AA\  to 7007\AA\  was used for these
with a grid spacing of $\delta\log_{10}(\lambda/\rm\AA)=1.8\times 10^{-4}$.

The template spectra were continuum subtracted and normalized prior to
orthogonalization. For simplicity, a constant variance and
unit weights were assumed at all wavelengths. The 2dF spectra 
were fluxed using an approximate mean
2dF response curve derived from photometric standards. 
Continuum subtraction was done at each point by
subtracting the local median calculated over a 100-bin wide window
centered at that 
point. The continuum subtracted spectra were normalized so that the sum
of the squares of fluxes in the continuum subtracted spectrum was unity.
With this normalization and unit weights the first term
on the right hand side of  Equation~\ref{newchi2} equals one, leading
to a particularly simple expression for $\chi^2$.

The resulting normalized spectra were orthogonalized using a standard
singular value decomposition routine.  A selection of the
resulting eigenspectra are shown in Figures \ref{figkes} and 
\ref{fig2dfes}.

The five eigenfunctions derived from the Kennicutt spectra are shown in
Figure~\ref{figkes}. Figure~\ref{fig2dfes} shows the 2dF eigenspectra
with the five highest eigenvalues. Orthogonalization of 91 
spectra leads to 91 eigenspectra but as discussed in section 2 
many of these represent noise. Five 2dF eigentemplates were retained
for the redshift determination and three Kennicutt eigentemplates. It can
be seen from Figure~\ref{figkes} that the first two eigenfunctions, derived
from the different data sets, are very similar. These two account for $>80$ \% of the
variation in the input data. The higher order eigenfunctions come out differently
for the different data sets, which is to be expected given the effect of noise
on the exact location of the Principal Components.

\subsection{Redshift Determination}

Redshifts were calculated using both the Kennicutt eigenspectra 
shown in Figure~\ref{figkes} and the 2dF eigenspectra shown
in Figure~\ref{fig2dfes}. As most spectral information is contained 
in the highest eigenvalue eigenspectra only the first three 
Kennicutt eigenspectra and the first five 2dF eigenspectra were 
retained. 

The 2dF spectra showed a number of residual sky features in the 
regions of strong atmospheric emission and absorption lines. Where these
are the strongest features in the spectrum there is a danger that
the correlation between the strong peaks in the eigenspectra 
(particularly the strong H$\alpha$ line) and the sky residuals
will be greater than the correlation between the templates and the
much weaker galaxy spectrum. As a preprocessing step before orthogonalization
sky residuals were removed in 60\AA\ bands around 5577, 5892, 
6300, 6363 and 7610\AA. The missing spectral bands were
interpolated using least squares fit to the spectrum on either
side and the spectra were rebinned onto the same wavelength grid
as the eigenspectra.

The rebinned spectra were  continuum subtracted and normalized in the 
same way as the template spectra. The expansion coefficients,
$b_j(z)$, can be quickly and efficiently found using 
fast fourier transforms. The FFT algorithms are most efficient 
if the total length of the series, $N$, is equal to a power of 2. 
In addition, because the
FFT treats the series as a periodic function of period $N$, $N$
must be greater than the sum of the length of the galactic and
template spectra to avoid errors in the cross-correlation calculation.
To this end, both template and galactic spectra were zero-padded
to the power of 2 greater than the sum of their lengths.

To illustrate the procedure, Figure~\ref{fig6sp}
shows a selection of six of the input spectra. They have been 
corrected for the sky residuals but not yet
continuum subtracted or normalized. The results are
discussed using the Kennicutt eigenspectra. Figure~\ref{figchi2}
shows the corresponding $\chi^2$ functions obtained.
Calculated and manual redshifts are
given in Table~\ref{tabz} along with the associated expansion
coefficients. 

Spectra (a) to (d) are typical of the 
majority of the spectra studied. They give calculated redshifts that
agree well with the manual redshifts.
The corresponding $\chi^2$ functions show  clear minima giving an
unambiguous determination of the redshift. Spectrum (e) is noisier and
had no manual redshift assigned to it previously, however, the $\chi^2$
function gives a clear, albeit weaker, peak at a redshift of 0.06. 
Spectrum (f) is the spectrum of a quasar included as a deliberate
outlier. The method clearly
fails to find a redshift for this spectrum, as expected since there
are no quasar spectra in the template set. There will always be
a minimum value of $\chi^2$ but it is clear from inspection
of the corresponding $\chi^2$ function that the associated redshift 
estimate is unreliable.

The failure of the method to find a redshift for the quasar
illustrates the importance of including all spectral types of
interest in the template set. The method will fail to find a
redshift for galaxies whose spectra differ fundamentally from
the template set (e.g. in instrumental resolution). 
However, in principal the method will work
for all spectral types, including emission line galaxies,
provided that the relevant spectral types lies within the
$m$-dimensional space spanned by the eigenspectra.
The power of the PCA method lies
in its ability to reduce the dimensionality of linear problems
from many templates to a few eigenspectra with no loss of
accuracy, and its resulting ability to filter out the noise from noisy
templates.

A side effect of this method is the ability to reconstruct `filtered'
versions of the spectra from the eigentemplates.
With only a few eigentemplates, the relative strength 
of the emission and absorption lines may not have fully converged, 
but a comparison of the reconstructed and original spectra 
helps to clarify how the method works. Figure~\ref{figrec}
shows the reconstructed spectra corresponding to the first five
original spectra shown in Figure~\ref{fig6sp}. No reconstruction is
given for the quasar spectrum.

Of course PCA will fail to reduce the problem space in non-linear
cases, a practical example might be if one has a sample of AGN with
very broad lines covering a large continuous range in velocity. However
for redshift work most galaxy spectra are unresolved, or only marginally
resolved in which case the variation can be accommodated in the
eigenspectra.

The ultimate  test of the method comes with a larger scale comparison of the
manually determined redshifts and the calculated redshifts. Figure~\ref{figzvz1}
shows the comparison between the manually determined redshifts and the
two sets of PCAZ redshifts calculated with the two sets of eigenspectra.
It is clear that the agreement between the PCAZ redshifts and the
manually determined redshifts are very good for this field with
a greater than $>98$\% success rate.  The 2dF eigenfunctions performed the
best giving only one mismatch at a redshift of 0.23. Clearly we need 
somewhat more than 104 spectra to determine the error rate at this 
high level of success --- something like 1000--2000 spectra are needed. 
We will look at this in more detail in Paper II.

Poor sky subtraction remain possible sources of error in the
automatic redshift determination. The PCAZ method took less than two minutes
of computer time to calculate the 104 redshifts. The measured 
scatter of the points on
the line is $\Delta z \leq 0.0005$ which is what we expect from the instrumental
resolution. 

With the PCAZ code it is trivial to turn off the steps of orthogonalization
and quadratic combination of cross-correlation functions --- this enables
us to reproduce the results of simple CCF analysis with the same template set.
This is also shown in 
Figure~\ref{figzvz1}, where the Kennicutt template with the highest CCF
peak gives the CCF redshift. It can be seen that for these high signal/noise
spectra the results are similar whether or not the templates are
diagonalized. This simply reflects the excellent quality of the 2dF spectra
with highly significant features for the algorithms to select.
We anticipated the PCAZ method would perform better than simple
CCF for lower signal/noise spectra (much of the initial testing was done
with such spectra before we had access to 2dF data). To demonstrate this
we add artificial gaussian noise to the 2dF data, both data
{\em and\/} templates and decrease the
continuum signal/noise by a factor of 3, so the galaxies
are typically $S/N=3$--10, and repeat
our analyses. Rerunning the PCA analysis gives virtually identical
2dF eigenfunctions, the redshift results
are shown in Figure~\ref{figzvz2}. It is evident
that PCAZ still performs at the 98\% level while the CCF method has
dropped to 93\% success rate. 
%
% NEW!!
% 
These results were obtained with minimal manual intervention and
illustrates that PCAZ is more robust in the low signal/noise regime.

Figure~\ref{figpcvcc} shows four of the spectra where the non-orthogonal cross
correlation method fails, labeled (a) to (d) in Figure~\ref{figzvz2}.
The lower curve in each panel shows the original 2dF spectrum. The
top curve shows the spectrum plus added noise. The central curve is
the PCA reconstruction of the noisy spectrum. Figure~9 
shows the corresponding cross-correlation functions and $\chi^2$
functions for the four spectra. The PCAZ results were
calculated using the first three Kennicutt eigenfunctions derived
from the same 5 Kennicutt spectra used for the non-orthogonal cross-correlation
method. For the spectra (a) to (c) the PCAZ method correctly locates
the redshift of the noisy spectra. The corresponding 
cross-correlation functions clearly show a peak at the same redshift, but
the noise peaks are as large. PCAZ simultaneously uses many templates, 
effectively averaging over the CCF noise. 

The fourth spectrum shows a case where both methods fail. The correct 
redshift is 0.115 but the presence of a sharp noise spikes, especially at
3950\AA\ has introduced   spurious correlations.

As surveys progress to thousands and tens of thousands
of galaxies we expect this relative advantage to increase: the derived
eigenfunctions will include more subtle natural variations in the range
of galaxy spectral features and will average over larger numbers of galaxies.

\section{CONCLUSIONS}

A new method of automatic redshift determination has been 
developed and shown to be capable of reproducing manually
determined redshifts with a minimal amount of manual 
intervention. The method is a superior generalization of
cross-correlation and has the potential to provide a 
sounder mathematical basis for confidence in the final
redshifts. The expansion coefficients generated can
be used to reconstruct noise filtered versions of the spectra
and have the potential to be used for a basic classification
of the spectra. The method proves more robust in the low signal/noise
regime than independent cross-correlation
and has greater potential for very high success rates
in upcoming very large redshifts surveys.

This concludes the introduction and illustration of the mathematical
principles behind PCAZ. 
In Paper II in this series we will be looking in more detail at
the reliability and the robustness of the method with much larger data sets and 
we will consider in detail the treatment of the data with realistic errors,
the robustness with signal/noise and compute typical probability distributions
for redshift errors from PCAZ. We will also examine, via simulations, how this 
affects the measurement of derived bulk galaxy properties from very
large redshift surveys such as $P(K)$ and the galaxy luminosity function.

\acknowledgements

The authors would like to thank Ofer Lahav and Joss
Bland-Hawthorn for useful discussions and Matthew
Colless for providing half the manual redshifts. We are grateful 
to the 2dF Galaxy Survey Team for allowing us to use their early
data for this paper and to the 2dF Commissioning Team at the Anglo-Australian
Telescope for all their hard work. KGB acknowledges the Anglo-Australian
Observatory for their generous support of research for staff astronomers.
KD was supported by a AAO Summer Studentship. Computing facilities for this
research were provided by the AAO.

\figcaption[fig1.ps]{Eigenfunctions obtained using the Kennicutt spectra
listed in Table~1. The vertical scale
is the flux per unit wavelength in normalized units.
\label{figkes}}

\figcaption[fig2.ps]{First five eigenfunctions obtained using a
sample of 91 2dF spectra.
\label{fig2dfes}
}

\figcaption[fig3.ps]{The six 2dF spectra discussed in the text. The
spectra have been corrected for the sky residuals and divided by
the instrument response function.
\label{fig6sp}}

\figcaption[fig4.ps]{The $\chi^2$ functions (multiplied by a
constant variance) corresponding to the spectra
discussed in the text. Note the different vertical scales
on the 6 spectra.
\label{figchi2}}

\figcaption[fig5.ps]{noise filtered reconstructions of the  five spectra
(a) to (e). The spectra are reconstructed from the first three 
Kennicutt eigenfunctions.
\label{figrec}}

\figcaption[fig6.ps]{Comparison with manual redshifts in the
SGP463 2dF field for the three automated methods discussed in the text: 
(a) PCAZ
redshifts determined using the eigenfunctions derived from
the NGP359 2dF field, (b) PCAZ redshifts
determined using the eigenfunctions derived from the Kennicutt templates,
(c) simple cross-correlation with the Kennicutt templates, picking 
the best peak. 
\label{figzvz1}}

\figcaption[fig7.ps]{As Figure~\protect\ref{figzvz1}, this time with the continuum
signal/noise degraded to the range $3$--10 for both the test data (NGP351)
and in the 2dF field (SPG463) used to construct the eigenfunctions in (a).
\label{figzvz2}}

%
% NEW!! -> Two new figures
%
 
\figcaption[fig8.ps]{Noise degraded spectra and reconstructions
corresponding to points a-d in Figure~\protect\ref{figzvz2}.
The top curve in each panel is the noise degraded input spectrum,
the second curve is the PCA reconstruction from the first three
Kennicutt eigenfunctions.
The lowest curve is the original 2dF spectrum from which the input
spectrum was derived.
\label{figpcvcc}}
 
\figcaption[fig9.ps]{ Cross-correlation functions from
a simple non-orthogonalized cross-correlation approach
and $\chi^2$ from PCAZ for
the spectra in Figure~\protect\ref{figpcvcc}. The five 
individual cross-correlation functions for each spectrum are
plotted on the same graph.}
\label{fig9}

\onecolumn

\begin{table}
\centering
\caption{Galactic spectra (Kennicutt (1992))
used to construct the first set of eigenspectra}
\label{kensp}
\bigskip\bigskip
\begin{tabular}{lc}
\hline
Galaxy & Morphology \\
NGC3379 & E0 \\
NGC4889 & E4 \\
NGC5248 & Sbc \\
NGC2276 & Sc \\
NGC4485 & Sm/Im \\
\hline
\end{tabular}
\end{table}

\begin{table}
\centering
\caption{Redshifts and Expansion coefficients for spectra (a) to (e)}
\label{tabz}
\bigskip\bigskip
\begin{tabular}{lccccc}
  & PCAZ & Visual Inspection & $b_1(z)$ & $b_2(z)$ & $b_3(z)$ \\
         & Redshift & Redshift &&& \\
a & 0.0674 & 0.0676 &  1.193 &  0.025 &  0.029  \\
b & 0.1411 & 0.1412 &  0.075 &  0.833 &  0.001  \\
c & 0.2379 & 0.2384 &  0.069 &  0.752 &  0.003  \\
d & 0.1809 & 0.1809 &  1.104 &  0.176 & -0.068  \\
e & 0.0600 &        &  0.683 &  0.139 &  0.003  \\
\end{tabular}
\end{table}

\clearpage

\input psfig
\psfig{file=figure1.ps,height=19.5cm}{\hfill \bf Glazebrook \etal\ Fig.~1}\newpage
\psfig{file=figure2.ps,height=19.5cm}{\hfill \bf Glazebrook \etal\ Fig.~2}\newpage
\psfig{file=figure3.ps,height=19.5cm}{\hfill \bf Glazebrook \etal\ Fig.~3}\newpage
\psfig{file=figure4.ps,height=19.5cm}{\hfill \bf Glazebrook \etal\ Fig.~4}\newpage
\psfig{file=figure5.ps,height=19.5cm}{\hfill \bf Glazebrook \etal\ Fig.~5}\newpage
\psfig{file=figure6.ps,height=19.5cm}{\hfill \bf Glazebrook \etal\ Fig.~6}\newpage
\psfig{file=figure7.ps,height=19.5cm}{\hfill \bf Glazebrook \etal\ Fig.~7}\newpage
\psfig{file=figure8.ps,width=\hsize}{\hfill \bf Glazebrook \etal\ Fig.~8}\newpage
\psfig{file=figure9.ps,height=20.5cm}{\hfill \bf Glazebrook \etal\ Fig.~9}\newpage

\label{lastpage}

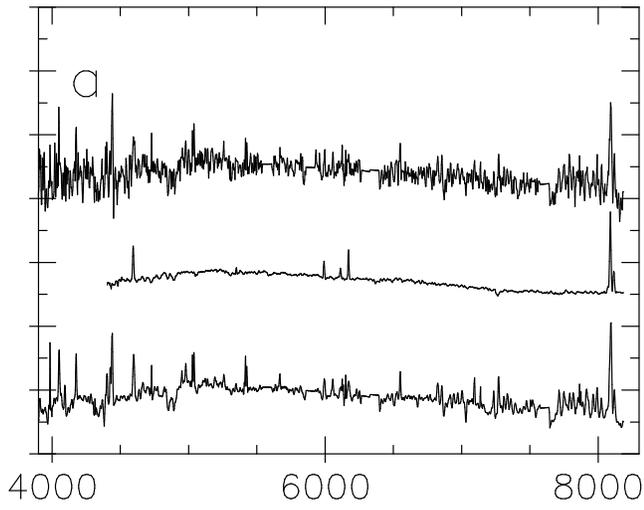
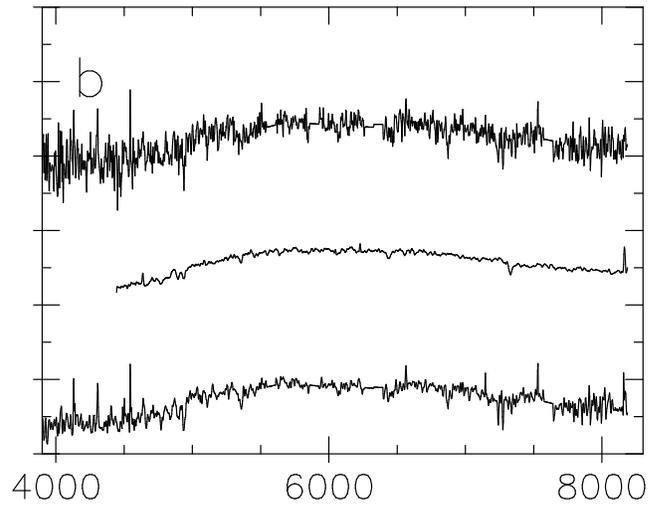
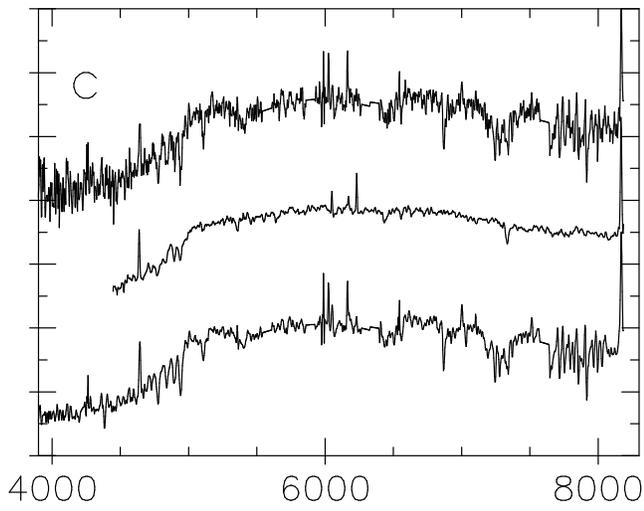
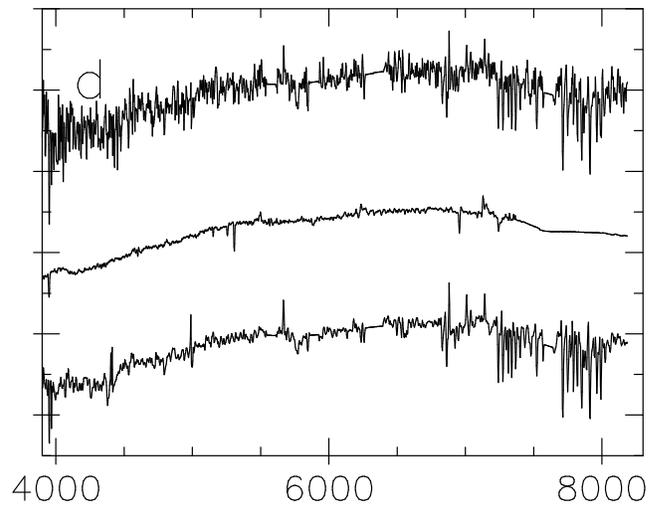
\end{document}